\documentclass{PoS}

\title{Strong coupling effective theory \\ with heavy fermions}

\ShortTitle{Strong coupling effective theory with heavy fermions}

\author{Michael Fromm, Jens Langelage, \speaker{Stefano Lottini}, Owe Philipsen\\
        Institut f\"ur Theoretische Physik - Johann Wolfgang Goethe-Universit\"at\\
        Max-von-Laue-Str. 1, 60438 Frankfurt am Main, Germany \\
        E-mail: \email{[last name]@th.physik.uni-frankfurt.de}}

\abstract{
	We extend the recently developed strong coupling, dimensionally reduced Polyakov-loop effective theory
	from finite-temperature pure Yang-Mills to include heavy fermions and nonzero chemical potential
	by means of a hopping parameter expansion.
	Numerical simulation is employed to investigate the weakening of the deconfinement transition as a function of the
	quark mass.
	The tractability of the sign problem in this model is exploited to locate the critical surface in the $(M/T, \mu/T, T)$ space
	over the whole range of chemical potentials from zero up to infinity.
}
\FullConference{XXIX International Symposium on Lattice Field Theory \\
                 July 10 -- 16 2011\\
                 Squaw Valley, Lake Tahoe, California}

\newcommand{\eq}{\begin{equation}}
\newcommand{\qe}{\end{equation}}
\newcommand{\hb}{\overline{h}}
\newcommand{\de}{\mathrm{d}\hspace{0.1em}}
\newcommand{\Real}{\mathrm{Re}}
\renewcommand{\det}{\mathrm{det}}
\newcommand{\Trace}{\mathrm{Tr}}
\newcommand{\avg}[1]{\langle #1 \rangle}
\newcommand{\FIGSIZE}{4.203cm}
\newcommand{\FIGSIZEROT}{4.21cm}
\begin{document}

\section{Introduction}
Solving QCD via lattice simulations
has proven to be a formidable problem, even unsurmountable, as of now, as soon as the
quark chemical potential $\mu = \mu_B/3$ is switched on, i.e.~if the finite-density part of the phase space is under study.
Hence, various alternative approaches have been developed to gain knowledge;
some are based on $\mu=0$ and extrapolate to finite $\mu$, while other rely on building effective models and approximate descriptions 
that capture, to a certain extent, the basic dynamics of the system under study.

A dimensionally-reduced effective theory based on strong-coupling expansion was introduced 
for the pure gauge sector in \cite{efftheory_puregauge}:
it offers robust predictive power in locating the thermal transition, can be improved order by order in a systematic fashion
and studied numerically with relatively small efforts.
Here we report on the inclusion of heavy fermions in the theory, implemented through a hopping-parameter expansion, and on
the introduction of a nonzero chemical potential, with a sign problem well under control even at large $\mu$.

This contribution offers a sketchy overview on the subject and focuses only on some of its features:
for a more detailed discussion, we refer the interested reader to \cite{efftheory2011}.

\section{Effective theory}
\label{sec:eft}
The theory under study comes from applying simultaneously strong-coupling and hopping parameter expansions to the Wilson action 
(on a lattice with temporal extent $a N_\tau = 1/T$, lattice spacing $a$, and gauge coupling $\beta$);
it is then suitable to investigate, with the advantages of a simplified, dimensionally reduced model,
the heavy quark region of the QCD phase space.

It is possible to integrate out the spatial links by means of a strong coupling expansion, which results in 
an action whose terms are each given by an effective coupling
(function of the original parameters $\beta$, $N_\tau$, $\mu$ and the hopping parameter $\kappa$)
and consist of Polyakov loops $L_i \equiv \Trace W_i = \Trace \prod_{\tau=1}^{N_\tau} U_0(i,\tau)$.
The partition function thus correctly reproduces the $Z_3$ centre symmetry of the gauge sector as well as its
breaking by the introduction of a finite quark mass $M$.
In practical applications, we restrict ourselves to just a few terms in such an effective action.

A remarkable aspect of this theory is that the definition of its partition function is not expressed with an action linear in the 
couplings; this is due to the possibility of performing a partial resummation among certain classes of graphs, which appears 
to improve convergence. 
Also, this calls for a careful definition of the suitable observables to characterise the phase structure.

\begin{figure}
 	\begin{center}
 	\includegraphics[height=\FIGSIZE]{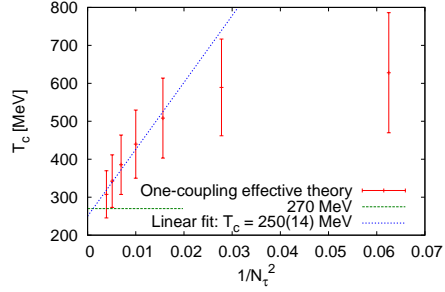}
 	\caption[]{Continuum limit of the pure gauge transition temperature $T_c$, from the effective theory.}
 	\label{fig:puregauge_continuum}
 	\end{center}
 \end{figure}

Moreover, in the pure gauge case it has been already observed that the resulting phase transition, albeit remaining first-order,
is much weaker than in the linear (i.e.~un-resummed) case, thus resembling QCD more closely. The critical effective coupling $\lambda_0$
for the quarkless theory can be translated into a table of $\beta_c(N_\tau)$ by means of strong-coupling mappings,
obtaining results close to those of full 4D simulations, allowing for a continuum extrapolation, 
which yields $T_c=250(14)$ MeV
(Fig.~\ref{fig:puregauge_continuum}).
 
Heavy quarks enter the model through an expansion in the hopping parameter $\kappa$; this results in a sum over closed 
loops which translates 
to a series expansion in mixed powers of $\kappa$ and $u$ (the latter being the first non-trivial coefficient in the
character expansion of the Yang-Mills action, $u(\beta)=\beta/18+\ldots$).
Partial resummations within classes of similar terms lead to writing the quark contribution in the form of a determinant.
If a quark chemical potential $\mu$ is turned on, each loop will pick up an additional factor \mbox{$e^{a\mu N_\tau}=e^{\mu/T}$}
raised to the power of its winding number.

When rewriting the model in terms of $L_i$, a ``potential'' term appears, encoding the reduced Haar measure on the group
and the Jacobian from expressing each $L_i$ as:
\eq
	\de L\,e^{V} = \de \theta \de \phi e^{2V}\;,
		\quad\;L(\theta,\phi)=e^{i\theta}+e^{i\phi}+e^{-i(\theta+\phi)}\;,\quad e^{2V(L)} = 27-18|L|^2+8\Real L^3 - |L|^4\;;
\qe
the partition function studied has then the form:
\eq
	Z(\lambda,h,\hb) = \int \prod_x \de L_x e^{V_x}
		\Bigg( \prod_{<ij>} \Big[ 1 + 2 \lambda \Real L_iL^*_j \Big] \Bigg)
		\Bigg( \prod_x \det \Big[ (1+hW_x)^{2N_f} (1+\hb W_x^\dagger)^{2N_f} \Big] \Bigg)\;,
	\label{eq:partitionfunction}
\qe
with effective couplings $\lambda(\beta,N_\tau,\kappa)$ as given in \cite{efftheory_puregauge, efftheory2011},
and, to leading order, $h=(2\kappa e^{a\mu})^{N_\tau}$ and $\hb=(2\kappa e^{-a\mu})^{N_\tau}$ (expressed to higher orders in \cite{efftheory2011}).
The number of flavours $N_f$ is from now on set to one, although thanks to the small values of $h,\hb$ involved
a linear approximation can be safely used to restore $h\to N_f h$.

One can express the fermion part entirely as a function
of $L, L^*$:
\eq
	Q_x \equiv \det[(1+hW_x)(1+\hb W_x^\dagger)]^2 = [(1+hL_x+h^2L_x^*+h^3)(1+\hb L_x^*+\hb^2 L_x + \hb^3)]^2\;\;.
\qe

\begin{figure}
	\begin{center}
	\raisebox{0.2cm}{\includegraphics[height=3.6cm]{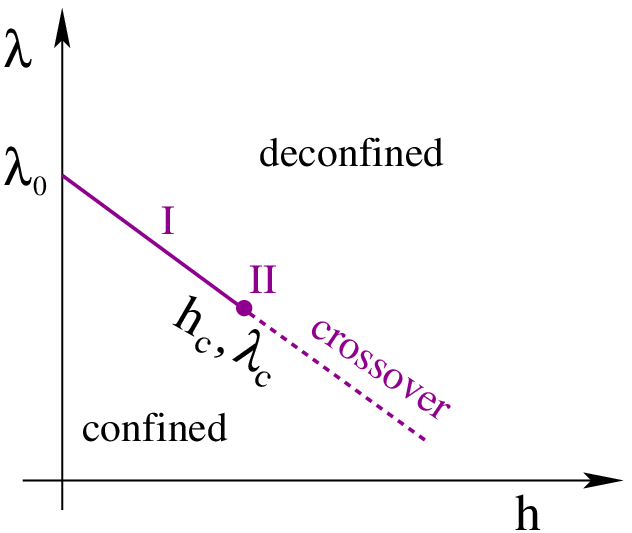}}
	\includegraphics[height=3.9cm]{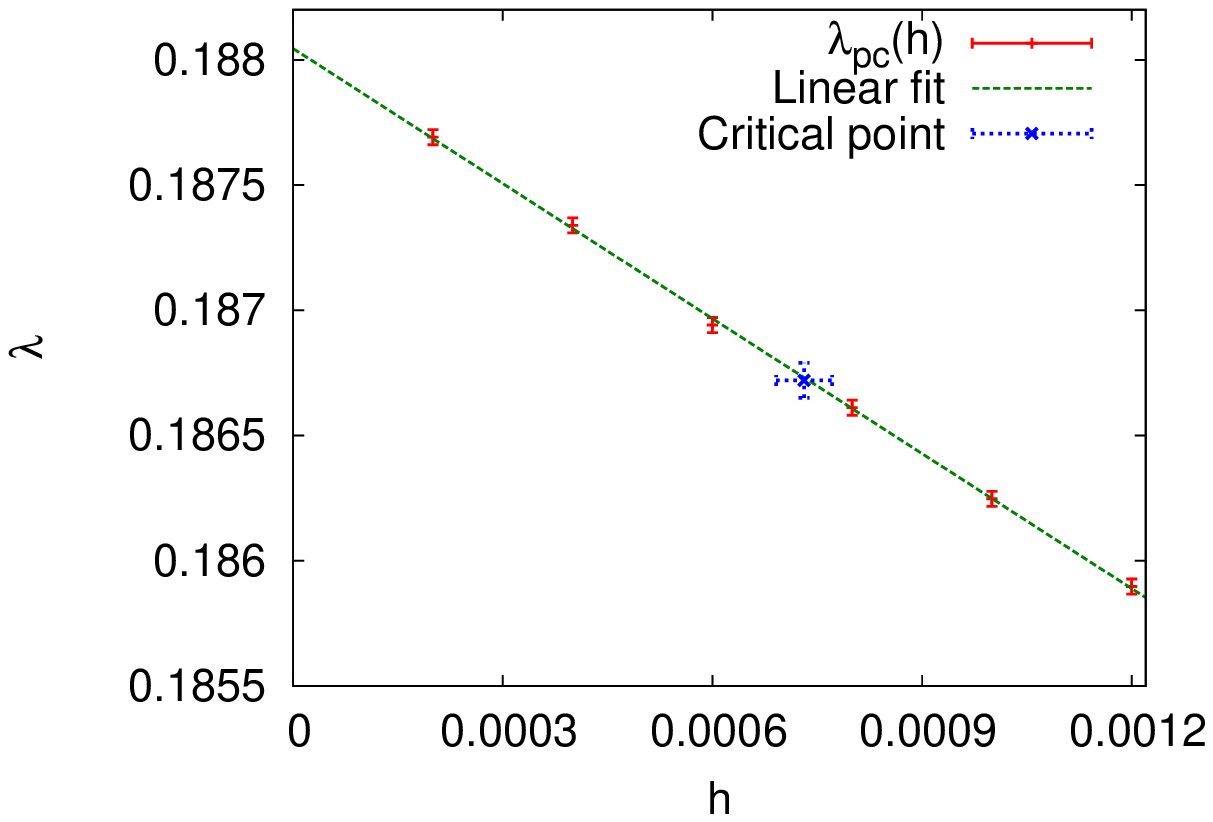}
	\caption[]{Left: Expected phase space for zero chemical potential. Right: Pseudocritical line for $\mu=0$.
	The line is a fit to Eq.~\ref{eq:muzero_critline}. Also shown is the critical point.}
	\label{fig:muzero_phase_space}
	\end{center}
\end{figure}

Nonlinearities aside, the above partition function can be compared to a three-state Potts model with a spin-spin interaction ($\sim \lambda$)
and an external magnetic field ($\sim h, \hb$) acting on each spin: from knowledge of the Potts case \cite{potts, potts_kim}, that has the same symmetry pattern, 
we expect a phase structure in $(h,\lambda)$ at zero chemical potential (meaning $h=\hb$) as depicted in Fig.~\ref{fig:muzero_phase_space} (left).
If the chemical potential is switched on, we have $h\neq\hb$, but the qualitative shape of the phase structure should not change.
In the latter case, however, we use for convenience the ``reduced'' $he^{-\mu/T} \equiv \tilde{h}$.

\section{Numerical results}
\label{sec:res}
The numerical investigation was performed with a Metropolis algorithm directly implementing the partition function
Eq.~\ref{eq:partitionfunction}: the sign problem appearing as soon as $\mu\neq 0$ is treated in the usual
way by folding the phase into the observable and updating according to the norm of the configuration weight.
It turns out that on systems as large as $N_s^3 = 24^3$ the average sign is well larger than zero up to
values of $\mu/T$ of three or more, thus posing no big trouble.\footnote{Also, the configurations yielding a minus sign
from the gauge part are extremely rare and can be in fact ignored at these system volumes and in the region
of parameter space of interest}
Each datapoint produced represents a statistics 
of about $10^6$ configurations, analysed with the binning technique in order to estimate uncertainties meaningfully.

Besides the usual observables, suitable for an action linear in its couplings, we also use their ``nonlinear''
counterparts (apart from trivial factors, they reduce to the former
for $\lambda,h,\hb\to 0$):
\eq
	E_\mathrm{lin} \equiv \frac{1}{3N_s^3}\sum_{<ij>}2\Real L_iL^*_j \;\;,\;
	Q_\mathrm{lin} \equiv \frac{1}{N_s^3} \Big| \sum_i L_i \Big| \;\;;\;
	E \equiv \frac{1}{\lambda} \frac{1}{3N_s^3}\sum_{<ij>} \log \Big( 1 + 2\lambda \Real L_iL_j^* \Big)\;\;,\;
	Q \equiv \frac{1}{h}\frac{1}{N_s^3}\sum_x \log Q_x\;\;.
	\label{eq:basic_observables_definition}
\qe
From these observables, the susceptibility and the Binder fourth cumulant have been built as 
\mbox{$\chi_O = N_s^3 (\avg{O^2}-\avg{O}^2)$} and $B_{4,O} = \frac{\avg{(O-\avg{O})^4}}{\avg{(O-\avg{O})^2}^2}$.
The main goal of this work is to map the phase structure in the $(\tilde{h},\lambda,\frac{\mu}{T})$ space: first, the case of zero chemical
potential is studied, then we introduce a real $\mu$.

\subsection{Zero chemical potential}
The investigation proceeds in two steps: first, the pseudo-critical line $\lambda_{pc}(h)$ is mapped,
subsequently its critical point $(\lambda_c,h_c)$ is located.
The pseudo-critical line is found by fixing six values of $0.0002\leq h \leq 0.0012$, and for each 
value by performing a $\lambda$-scan at various system volumes, identifying four
volume-dependent pseudocriticality estimators (extrema of susceptibility and Binder cumulant of $E_\mathrm{lin},Q_\mathrm{lin}$).
Then, for each of those estimators, an infinite-volume extrapolation
\mbox{$\lambda_{pc}(h,N_s) = \lambda_{pc}(h) + c_1(h) N_s^{-\alpha}$}
gives a thermodynamic limit which we find to mutually agree.
The whole pseudocritical line is parametrised as
\eq
	\lambda_{pc}(h) = \lambda_0 - a_1 h \;\;.
	\label{eq:muzero_critline}
\qe
A fit to the six points works well (Fig.~\ref{fig:muzero_phase_space}, right) and gives $a_1=1.797(18)$ and $\lambda_0=0.18805(1)$,
the latter roughly reproducing 
the pure-gauge critical point found in \cite{efftheory_puregauge}.
\footnote{The slight deviation between the two results is due to the small volumes used in \cite{efftheory_puregauge} for the determination.}
The linearity can be explained by a first-order expansion, in the small couplings involved, of the free energy of the system
\cite{potts}.

\begin{figure}
	\begin{center}
	\includegraphics[height=\FIGSIZE]{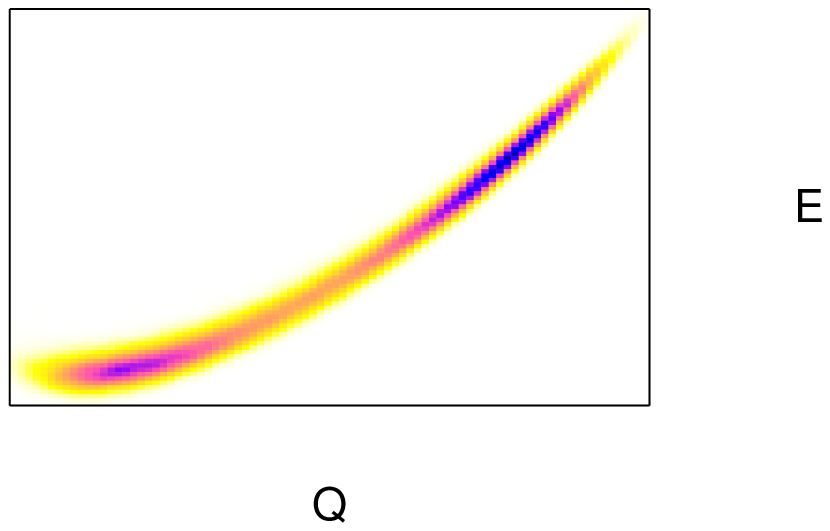}\qquad
	\includegraphics[height=\FIGSIZEROT]{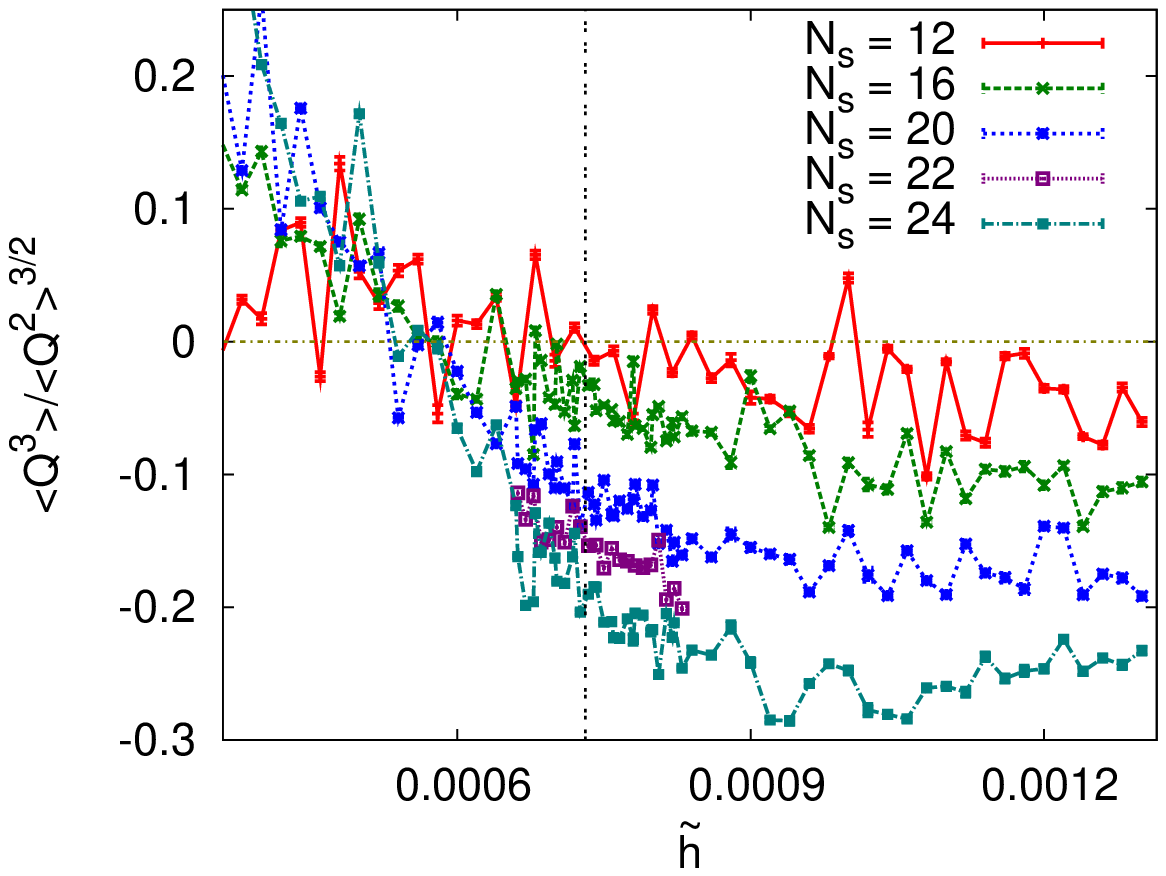}
	\\
	\includegraphics[height=\FIGSIZE]{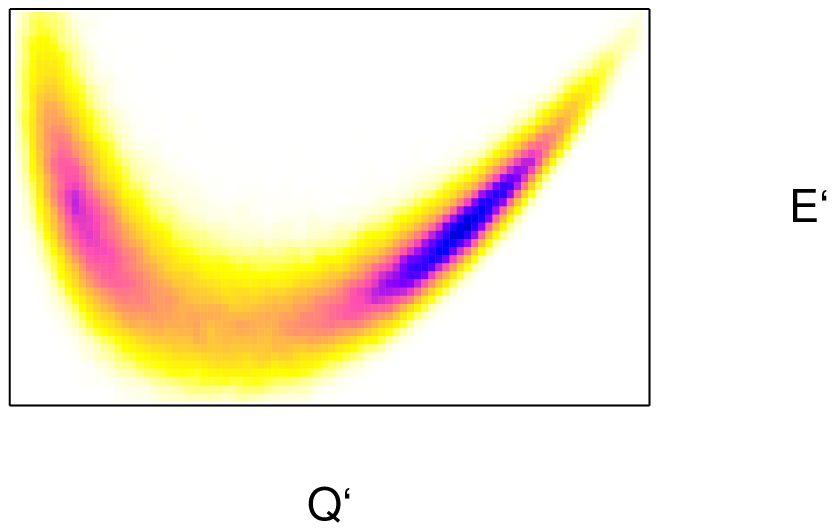}\qquad
	\includegraphics[height=\FIGSIZEROT]{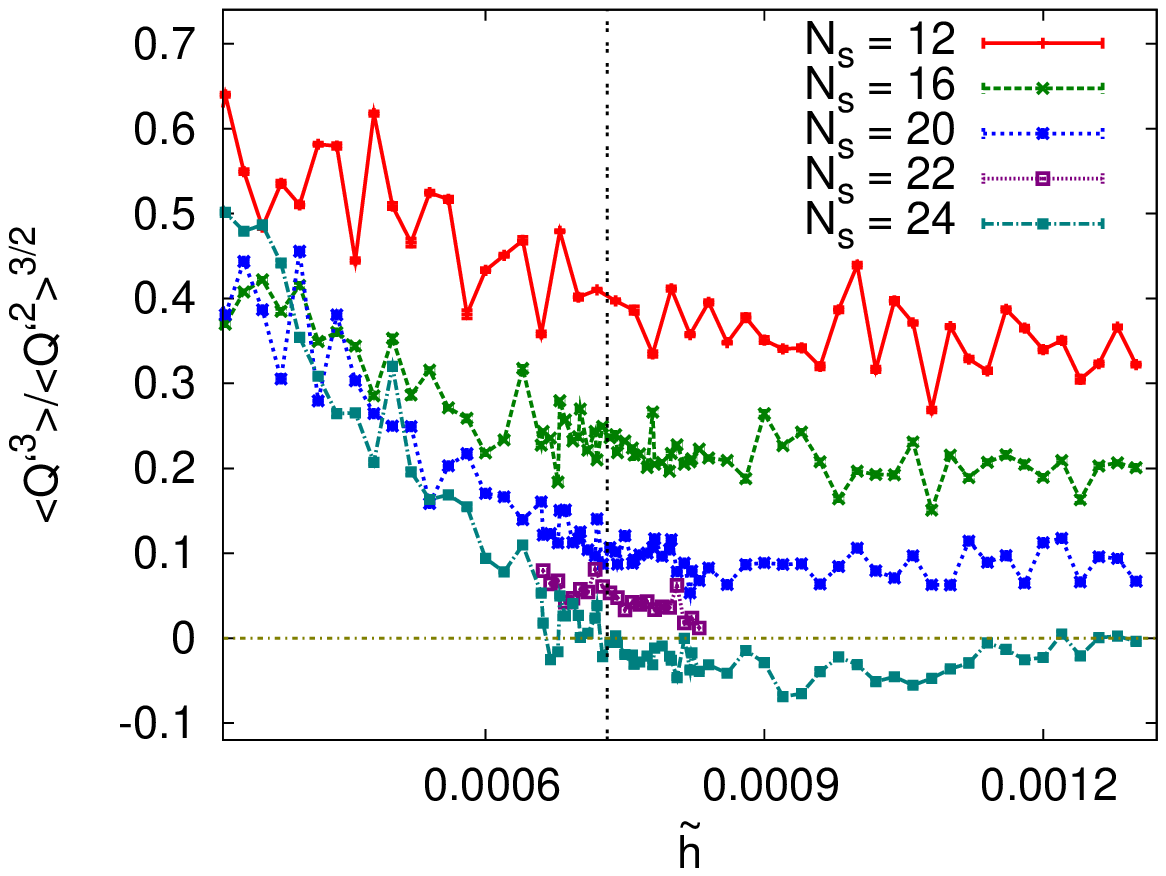}
	\caption[]{Effect of the rotation $(E,Q)\to(E',Q')$ in locating the $\mu=0$ critical line. Top panel: original
	observables. Bottom panel: rotated observables. On the left the (normalised, centred) histogram is shown for the sample case $N_s=20$,
	$h = 0.000742$, on the right the behaviour of the third moment of $Q$ ($Q'$) is plotted for various system volumes.
	Vertical lines mark the critical $h_c$. Note that the largest volumes show, for the rotated $Q'$, a third moment 
	essentially zero around the critical point.}
	\label{fig:rotation}
	\end{center}
\end{figure}

In order to locate the critical point, we switch to the nonlinear observables, Eq.~\ref{eq:basic_observables_definition},
and measure them along $\lambda_{pc}(h)$ at various system sizes up to $N_s=24$.
Close to the critical point, and focusing on $Q$, we expect the following scaling laws for the susceptibility and Binder cumulant:
\eq
	\chi_Q = N_s^{\gamma/\nu} f_{\chi_Q}(x)\;\;,\;\;B_{4,Q} = f_{B_{4,Q}}(x)\;\;;\;\;x\equiv (h-h_c)N_s^{1/\nu}\;,
\qe
with critical indices dictated by the three-dimensional Ising universality class, i.e.~$\gamma/\nu\simeq 1.962, \nu\simeq 0.6302$.
Moreover, universality also implies $f_{B_{4,Q}}(0)\simeq 1.604$.
Writing $f(x)$ as a series in $x$, the susceptibility and Binder cumulant data were fitted to the above expectation
keeping $N_s\geq 20$, with rather stable results against different truncations for $f(x)$, different scaling windows, fixing
or leaving free the critical indices, and we get the final values \mbox{$(\lambda_c, h_c) = \Big( 0.18672(7), 0.000731(40) \Big)$}.

Another, more rigorous method is employed in \cite{rotation_rummukainen} to identify the critical point (see also \cite{rotation_2} for an application
to a model similar to ours).
The two-dimensional distribution
of $(E,Q)$ is subject to a rotation $ \to (E',Q')$, with zero covariance; then, the critical line is defined as the locus
where, in the thermodynamic limit, the third moment of the centred marginal distribution of $Q'$ vanishes, \mbox{$\frac{\avg{Q'^3}}{\avg{Q'^2}^{\frac{3}{2}}} = 0$}.
We explicitly verified that, around the critical point and for large enough volumes, the rotated $Q'$ essentially satisfies this requirement along the 
line identified as described above (Fig.~\ref{fig:rotation}).

\subsection{Real chemical potential}
As already observed, with the sign problem well under control for our purposes, we basically repeat
the analysis performed for $\mu=0$ at several values of $\mu/T$ up to $3.0$. The only difference is that, for each chemical potential,
we generate data only at a single point $(\lambda,\tilde{h})$
and then reweight all results to a whole 2D grid of points (the reweighting factors are complicated
by the nonlinear formulation Eq.~\ref{eq:partitionfunction}, but if one knows the target couplings in advance
the table of weights can be prepared as the configurations are explored by the Monte Carlo).

\begin{figure}
	\begin{center}
	\includegraphics[height=\FIGSIZE]{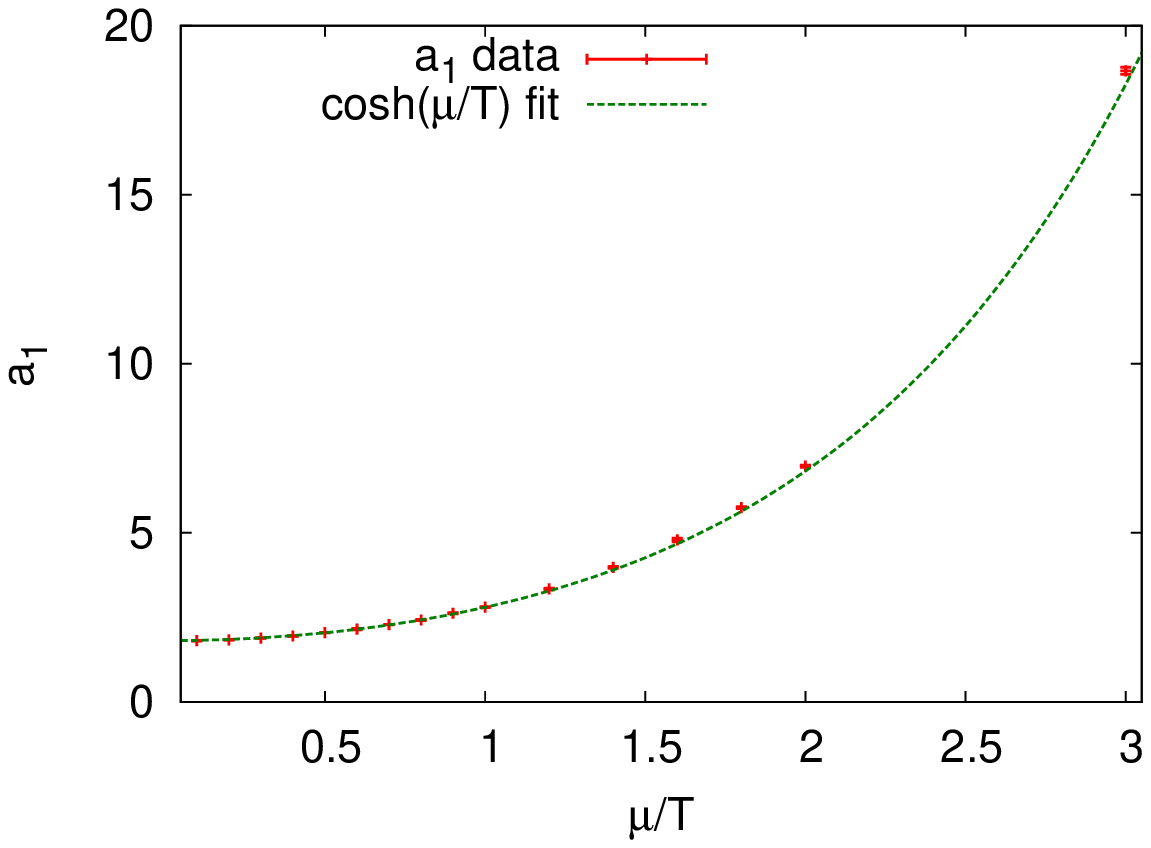}
	\includegraphics[height=\FIGSIZE]{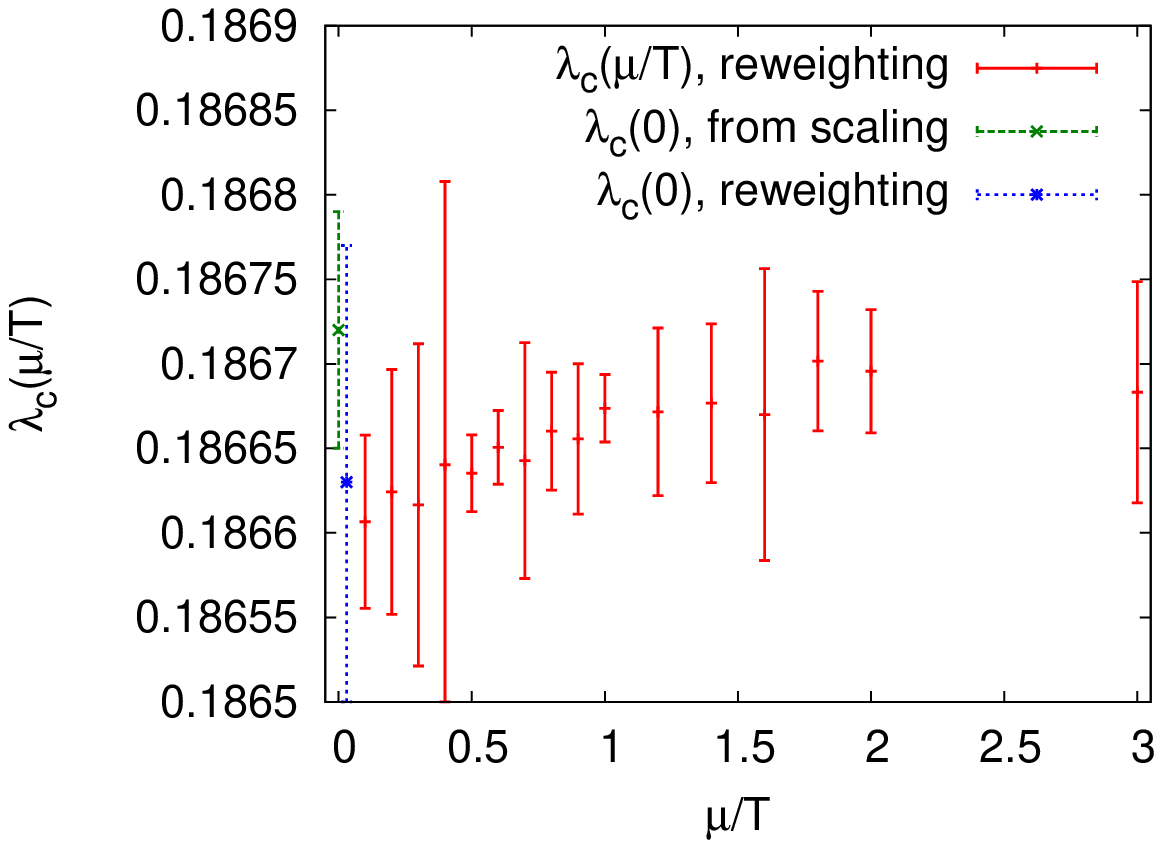}
	\caption[]{Left: slope $a_1$ of the pseudocritical line as a function of $\mu/T$, with its $\cosh(\mu/T)$ description.
	Right: the very weak dependence of $\lambda_c$ on the chemical potential. The point labelled ``$\lambda_c(0)$, reweighting''
	was found with the same technique as for finite $\mu$ as a cross-check.
	The slight drift in $\lambda_c$ is explained by a finite-size correction whose amplitude decreases as $\mu$ grows: indeed, the
	more accurate determination (``scaling'') for zero chemical potential and the large-$\mu$ values agree very well.}
	\label{fig:realmu_curves}
	\end{center}
\end{figure}

With the same statistics as for $\mu=0$, for each chemical potential the 2d grid of values $B_{4,Q}(\lambda,\tilde{h})$
was scanned for the line of local minima: the largest-volume result was taken as the pseudocritical line and fitted to
$\lambda_{pc}(h; \mu/T) = \lambda_0(\mu/T) - a_1(\mu/T)\tilde{h}$; again, a linear relation was sufficient,
and $\lambda_0$ turned out to be a constant compatible with the one in Eq.~\ref{eq:muzero_critline}.
Furthermore, it can be argued that, neglecting higher-order corrections, the slope of the curve depends on $\mu/T$ as
\mbox{$a_1(\mu/T) = C \cosh(\mu/T)$},
a behaviour that was confirmed numerically with $C = 1.814(3)$, in agreement with the $\mu=0$ slope (Fig.~\ref{fig:realmu_curves}, left).

\begin{figure}
	\begin{center}
	\includegraphics[height=3.75cm]{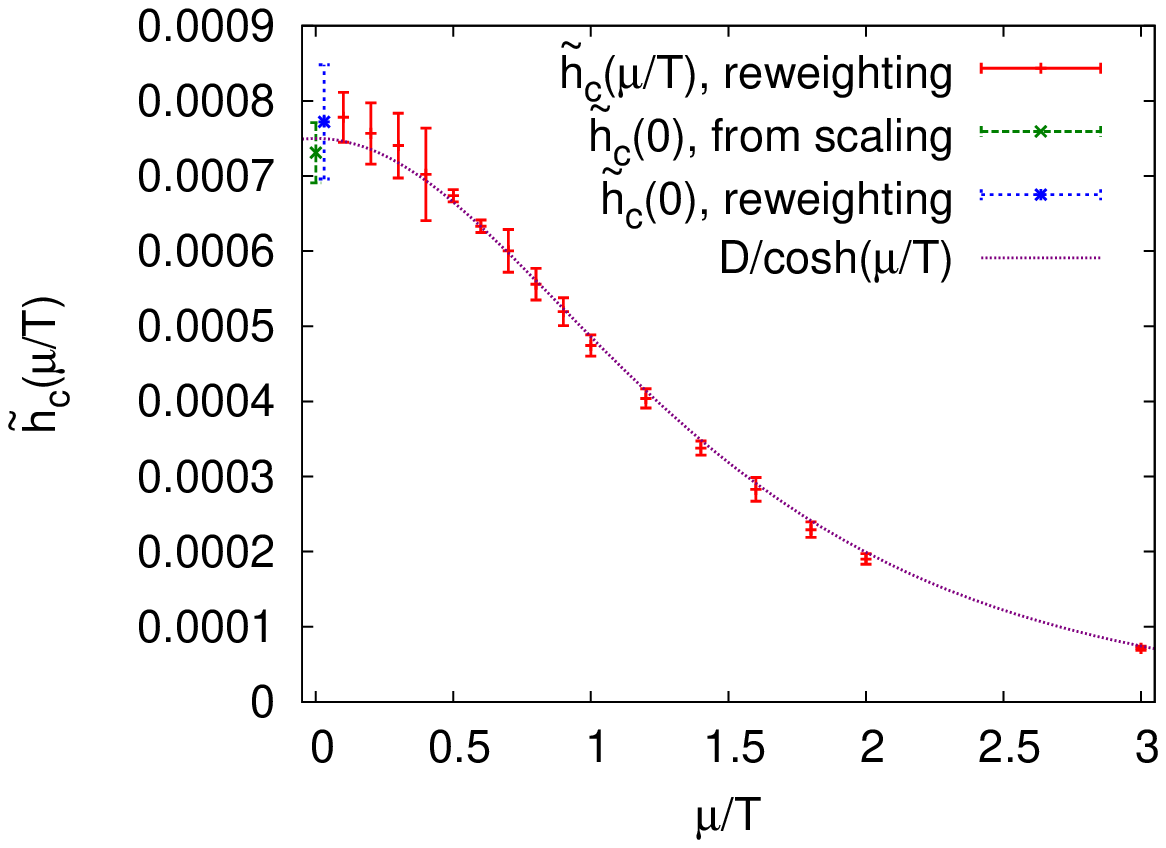}
	\includegraphics[height=3.75cm]{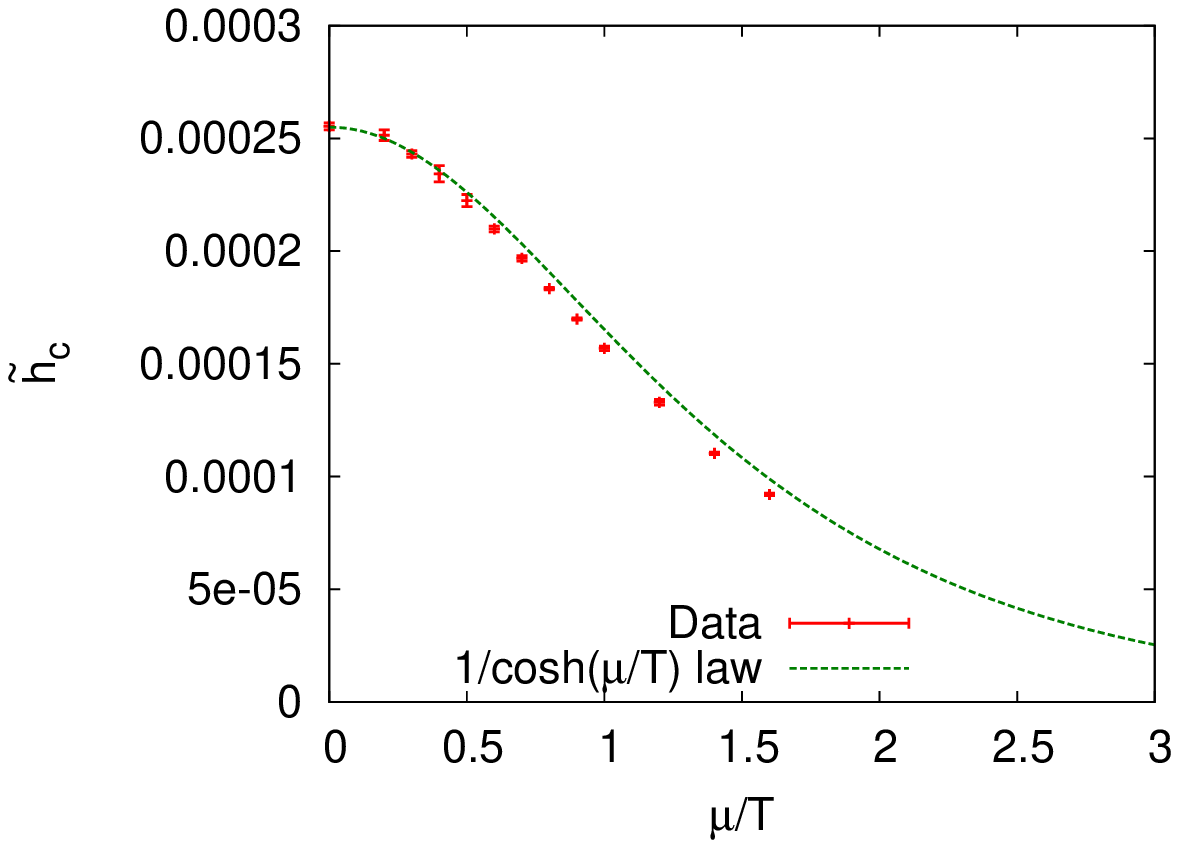}
	\caption[]{Left: the critical curve $\tilde{h}_c(\mu/T)$ along with the best-fit to Eq.~\ref{eq:cosh_htildecrit} for our effective model.
	Right: the equivalent figure in the case of the three-state Potts model \cite{potts_kim}: the fit curve, restricted to $\mu/T \lesssim 0.5$,
	shows that the same phenomenon of large-$\mu$ overestimation of the critical $\tilde{h}$ occurs in both models as a (tiny) deviation
	from the assumption of constant $\lambda_c(\mu/T)$. We plotted the data in terms of $\tilde{h}$ for ease of comparison.}
	\label{fig:final_curves_comparison}
	\end{center}
\end{figure}

As for the critical point determination, the $\mu=0$ fits encouraged us to define $\tilde{h}_c(\mu/T)$ as the value where
$B_{4,Q}=1.604$, with an uncertainty estimated from the difference between this definition and the $\tilde{h}$ at which the $B_{4,Q}$ for
the volumes $22^3$ and $24^3$ cross each other.
Remarkably, the critical $\lambda_c(\mu/T)$ shows little or no dependence on the chemical potential
(Fig.~\ref{fig:realmu_curves}, right), which allows to rewrite the parametrisation of $a_1(\mu/T)$ as
\eq
	\tilde{h}_c(\mu/T) = \frac{D}{\cosh(\mu/T)}\;\;.
	\label{eq:cosh_htildecrit}
\qe
A fit of the measured points to the above curve works indeed well, giving $D=0.00075(1)$ in full agreement with 
the $\mu=0$ result (Fig.~\ref{fig:final_curves_comparison}, left).
There is, however, a slight deviation from the above law, that we ascribe to higher terms of the expansion in $(\mu/T)^2$
which, to first order, led to parametrising $a_1(\mu/T)$. Remarkably, a similar phenomenon occurs in the simpler case of an effective theory
constructed with the three-state Potts model, as can be seen in Fig.~\ref{fig:final_curves_comparison}, right.

\section{Conclusions and outlook}
\label{sec:con}
With the knowledge of the curve $\tilde{h}(\mu/T)$ we can use the heavy-quark approximate relation $\tilde{h} = N_f \exp(-M/T)$
and obtain the critical surface in the upper-right corner of the Columbia plot; in particular, for the sake of comparison with 
existing literature \cite{rotation_2, saito},
we quote here the $\mu=0$ values of $M_c/T$ and $\kappa_c(N_\tau=4)$ for $N_f=1,2,3$ respectively:
\eq
	\frac{M_c}{T} = \{7.22(5), 7.91(5), 8.32(5)\}\;,\;\kappa_c(N_\tau=4) = \{ 0.0822(11), 0.0691(9), 0.0625(9) \}\;.
\qe
By carrying on the expansions to higher orders, a more careful analysis of the feasibility of a continuum limit can be performed;
this program is indeed illustrated in \cite{efftheory2011}, along with the study of the imaginary-$\mu$ side of the phase space.
Another interesting direction of this study is the investigation of the low-temperature, large density limit of the theory, besides, of course,
the attempt to lower the fermion mass as much as possible, within the range of applicability of the hopping expansion.


\begin{thebibliography}{99}

	\bibitem{efftheory_puregauge}
  J.~Langelage, S.~Lottini and O.~Philipsen,
  JHEP {\bf 1102} (2011) 057
  [Erratum-ibid.\  {\bf 1107} (2011) 014]
  [arXiv:1010.0951 [hep-lat]];
  PoS {\bf LATTICE2010 } (2010)  196 
  [arXiv:1011.0095 [hep-lat]];

	\bibitem{efftheory_proceeding}
	S.~Lottini, O.~Philipsen, J.~Langelage,
	Acta Physica Polonica B Proc.~Suppl.~4, No.~4 [2011] 721 
	[arXiv:1105.5284 [hep-lat]].

	\bibitem{rotation_rummukainen}
  K.~Rummukainen, M.~Tsypin, K.~Kajantie, M.~Laine, M. Shaposhnikov,
  Nucl.\ Phys.\  {\bf B532}, 283-314 (1998) 
  [arXiv:hep-lat/9805013v1].


	\bibitem{rotation_2}
  C.~Alexandrou, A.~Bori\,ci, A.~Feo, P.~de Forcrand, A.~Galli, F.~Jegerlehner, T.~Takaishi,
  Phys.\ Rev.\  {\bf D60}, 034504 (1999) 
  [hep-lat/9811028].
	
 \bibitem{saito}
  H.~Saito {\it et al.} [WHOT-QCD Collaboration].
  arXiv:1106.0974 [hep-lat].


	\bibitem{efftheory2011}
	M.~Fromm, J.~Langelage, S.~Lottini, O.~Philipsen,
	arXiv:1111.4953 [hep-lat].

	\bibitem{potts}
  M.~G.~Alford, S.~Chandrasekharan, J.~Cox, U.-J.~Wiese,
  Nucl.\ Phys.\  {\bf B602}, 61-86 (2001) 
  [hep-lat/0101012].

	\bibitem{potts_kim}
   S.~Kim, Ph.~de Forcrand, S.~Kratochvila, T.~Takaishi,
   PoS(LAT2005) 166,
   [hep-lat/0510069].


\end{thebibliography}
\end{document}